**Transportation Needs Assessment for Rural Communities: A Case of Pickens County, Alabama**

**Riffat Islam**
Graduate Research Assistant
Department of Civil, Construction & Environmental Engineering
The University of Alabama
3013 Cyber Hall, 248 Kirkbride Lane, Tuscaloosa, AL 35487-0288
Email: rislam1@crimson.ua.edu

**Olga A. Bredikhina**
Assistant Research Economist, Alabama Transportation Policy Research Center
The University of Alabama
3055 Cyber Hall, 248 Kirkbride Lane, Tuscaloosa, AL 35487-0288
Email: oabredikhina@ua.edu

**Muhammad Sami Irfan**
Department of Civil, Construction & Environmental Engineering
Graduate Research Assistant
The University of Alabama
3013 Cyber Hall, 248 Kirkbride Lane, Tuscaloosa, AL 35487-0288
Email: mirfan@crimson.ua.edu

**Khadiza Tul Jannat**
Instructor, Department of Communication & Modern Languages
Southeast Missouri State University
Academic Hall, One University Plaza, Cape Girardeau, MO 63701
Email: kjannat@semo.edu

**Steven Jones, Ph. D.**
Director, Transportation Policy Research Center
Deputy Director, Alabama Transportation Institute
James R. Cudworth Professor of Department of Civil, Construction & Environmental Engineering
The University of Alabama
3024 Cyber Hall, 248 Kirkbride Lane, Tuscaloosa, AL 35487-0288
Email: sjones@eng.ua.edu

*Word count: 5,023 words + 9 tables (250 words per table) = 7,273*
*Submitted: August 1, 2022*

## ABSTRACT

The lack of transportation options and limited public transit service in rural areas of the U.S. may negatively affect residents' quality of life by limiting the opportunities for employment, community engagement, education, and quality healthcare. This study used a mixed-method approach combining quantitative and qualitative research methods to conduct the transportation needs assessment in Pickens County, Alabama that examined the access to healthcare, jobs, and other relevant resources across all population categories in the area, including vulnerable populations, such as senior citizens, people with disabilities, and low-income households. Additionally, this study investigated public perceptions of, and opinions about H.E.L.P. Inc., an origin-to-destination demand-response bus service for Pickens County residents. The study's findings may be useful to stakeholders, policymakers, and transportation services providers in the rural communities of the U.S. working develop new approaches that would help meet the transportation needs of local communities.

**INTRODUCTION**

Mobility and transportation have a significant impact on the quality of life in America's rural communities. Effective transportation systems provide access to healthcare, education, training, jobs, shopping centers, etc. (1). At the same time, the lack of transportation options and limited public transit service may limit residents' employment, community engagement, and education opportunities as well as result in missed or delayed healthcare appointments and increased healthcare expenditures (2). Public transit is especially important for those residents that do not have a reliable access to personal vehicle such as older adults, people with disabilities, youth, low-income households, etc. (3). Despite the importance of mobility and access to transportation for the quality of life, many rural communities in the U.S. are lacking public transportation services (3). In the United States, the lack of access to transportation is reported as the third most commonly cited obstacle to receiving health services for older persons, preventing 3.6 million people from getting healthcare services (4). Numerous Temporary Assistance for Needy Families (TANF) beneficiaries need reliable public transit to aid their job search and daily commute. Rural areas pose transportation challenges due to low population densities and availability of nearby resources (5).

This study presents and discusses the results of a transportation needs assessment in Pickens County, Alabama that examined the access to healthcare, jobs, and other relevant resources across all population categories, including vulnerable populations, such as senior citizens, people with disabilities, and low-income households. The primary objective of this research is to improve the understanding of transportation-related choices and challenges experienced by residents of rural areas with low population densities like Pickens County, Alabama. Previous research studies that explored the transportation needs in rural communities in the U.S. primarily focused on the impact of mobility in rural communities on the access to healthcare and residents' outcomes and on the impact of public transit in rural communities on the quality of vulnerable populations such as the elderly and individuals with disabilities. Although the survey conducted as a part of this research study included vulnerable populations among the survey's participants, the study explores the access to transportation options and transportation-related challenges experienced by all residents of Pickens County and not only the vulnerable groups of the community.

Additionally, this research study investigates public perceptions of, and opinions about, H.E.L.P. Inc., an origin-to-destination demand-response bus service for Pickens County residents (6). Therefore, the qualitative findings about respondents' experiences with the H.E.L.P. bus and their suggestions regarding possible ways to improve the service may be helpful to stakeholders, policy-makers, and researchers interested in understanding and improving the experiences of passengers using demand-response transportation services in rural communities across the U.S.

This study employed a mixed-method research approach, incorporating quantitative and qualitative data collected from the paper-based survey and the focus group discussions, respectively. Both the survey and the focus group discussions were conducted among the residents of Pickens County, Alabama. The study's findings may be generalized to other rural communities across the U.S. and may be useful to stakeholders, policymakers, and transportation services providers in developing new approaches to meet the transportation needs of local communities.

## BACKGROUND

The U.S. Department of Transportation defines as "rural" any area outside of an agglomeration of 5,000 people for highway functional classification purposes and any area outside of an agglomeration of 50,000 for planning purposes (7), while the U.S. Census Bureau defines any agglomeration of 2,500 or more  as "urban" and everywhere else as "rural" (8). Rural communities compose 75% of the U.S.' land area and only 17% of the population (9). Rural areas pose unique challenges to transportation providers compared to their urban counterparts. Due to low population densities, limited financial and technical resources, and longer distances, rural transportation stakeholders and transportation services providers may need to develop and utilize a different set of approaches to help address transportation-related challenges in rural areas (10, 11). In the U.S., a high proportion of older adults and low-income individuals and families live in rural areas (12, 13). In fact, over 20% of older Americans live in rural areas and many of these adults are concentrated in states where more than half of older populations are in rural areas (13). Additionally, with the majority of low-income households living in rural areas and central parts of cities many low-income workers have challenges accessing jobs, training, and other essential services such as healthcare and childcare if these services are located outside of their areas of residence (12). At the same time, improving rural transportation options may not only improve residents' quality of life but also significantly benefit local communities. Improving the access to transportation in rural areas may help preserve and grow local economies by expanding the customer base for various services and businesses (e.g. medical facilities and shopping centers), raising employment, reducing government and local authorities' spending on programs such as unemployment compensation and food stamps, and helping families decrease their transportation costs (12)

Previous studies have shown that there is a lack of research about the transportation needs of older adults, low-income and at-risk populations and that understanding the local context of communities is important in order to ensure that residents' transportation-related needs are being met. Previous studies about transportation needs in rural areas have utilized statistical modeling and survey data to project and understand the demand for transportation in rural areas (14, 15).

A study completed by Arcury et al. in 2005 used a survey of households in 12 western counties of North Carolina to examine the association of transportation and health care utilization in rural areas in the state (14). The study found that having a personal driver's license doubled the number of chronic care and regular care visits while using public transportation slightly increased the number of chronic care and regular care visits per year (14).  A different study, conducted by Flethcher et al. in 2010 used three sequential studies including an in-depth longitudinal qualitative study, multiple-methods study, and a dual-frame household survey to better understand the transportation barriers facing low-income rural families and the association between transportation access and economic outcomes (15). The study generated multiple findings including the fact that those individuals that have to rely on others for a ride often see such transportation option as "unreliable", and that those workers that have a reliable access to transportation are more likely to have higher earnings. The study found a disconnect between the perspectives of community professionals and welfare recipients when it comes to transportation-related challenges in the community highlighted the importance of stakeholders and community members working together in order to effectively design strategies and approaches to address such challenges.

## METHODOLOGY
### Participants
For both the paper-based survey and the focus group discussions, the study's target population included primarily adult residents of Pickens County, Alabama. The paper-based survey included 84 participants (with the mean age of 55 with the sample standard deviation of 22 years). Additionally, 63 individuals (with the mean age of 66 with the sample standard deviation of 14.6 years) participated in eight focus group focus group discussion (FGD) sessions that were held at four different locations in Pickens County: Gordo, Aliceville, Reform, and Carrollton. Participation in the paper-based survey was voluntary and no reward for participation was offered to the participants. The participation in FGD was also voluntary; however, participants received lunch and a nominal payment to cover travel expenses.

### Procedures
This study was conducted as a part of the partnership between The University of Alabama - Pickens County with the goal to study transportation barriers in Pickens County, Alabama. Researchers collaborated with the Pickens County Family Resource Center to recruit participants for the paper-based survey and FGDs. Additionally, formal letters were sent to several churches in Pickens County asking for their help in disseminating the survey questionnaire and FGD recruitment flyers among their members. FDG recruitment flyers were also disseminated in senior centers and the Division of Child Support Services offices located in Pickens County. FGD recruitment flyers contained brief information about the goals, format time and location of FGDs as well as the information on how to sign up for FGDs.

The total of eight focus groups were conducted in four distinct locations in Pickens County (Aliceville (Aliceville Public Library), Gordo (Gordo City Hall), Carrolton (Carrolton Service Center), and Reform (Reform Senior Activity Center), two focus groups discussions per each location. In the disseminated recruitment flyers, residents of Pickens County were invited to sign up for FGDs using the provided phone number. For each FGD, there were a total of eleven open slots available. If more participants wanted to join, they were added to the waiting list. Additionally, if no open slots were available, those residents that were interested in participating, were offered the option to sign up for a FGD at another location if any open slots still remained there. Participants' phone numbers were recorded, and they were contacted one day before FGD as a reminder. Not all those individuals that signed up came to participate in FGDs. Each FGD lasted between 1-1.5 hours. During a focus group discussion, two audio recorders were used to record the entire conversation among the participants and the moderator. The moderator and co-moderator also took notes during the discussion for further convenience. These audio recordings were transcribed and documented separately for each individual group. Throughout the transcriptions, pseudo names were used to keep the participants' responses anonymous. The confidentiality of survey and FGD participants was protected by the Institutional Review Board (IRB) at the University of Alabama.

**Measures**

The paper-based survey questionnaire consisted of 23 multiple choice questions divided into three categories: (1) transportation options, (2) questions about residents' experiences with the H.E.L.P. bus service, and (3) demographic questions. Table 1 explains the type of information that was collected in each of the survey sections.

**Table 1. Summary of survey questions**

| Question Category | Information Collected |
|---|---|
| *Transportation Options* | Means of getting around, vehicle ownership, access to healthcare, travel plans, access to facilities, transportation related challenges etc. |
| *Residents' experiences with the H.E.L.P. Inc. bus service* | Familiarity with HELP bus, experience with HELP bus etc. |
| *Demographics* | Personal information (gender, age, race, etc.), household information (size, income, vehicle ownership, etc.) etc. |

The focus group discussion questionnaire consisted of 12 questions developed following the guidance for designing and conducting effective focus group interviews by Morgan and Krueger (16). FGD were built around the same question themes that were asked in the survey questionnaire with the majority the discussion focused on residents' access to facilities and their experiences with the H.E.L.P. bus. During FGDs, the moderator ensured that each participant had sufficient amount of time to respond to questions posed by the moderator and comments posed by other participants. Additionally, designated observers took notes and ensured that the moderator did not miss any participants' comments.

**Mixed method**

This study employed a mixed-method research (MMR) design as it combined qualitative and quantitative data collected from survey questionnaires and focus group discussions conducted among the residents of Pickens County. MMR is a research method frequently used in the fields of social sciences, psychology, education, and health sciences research (17-20). MMR integrates qualitative and quantitative approaches at the data collection, data analysis, and results interpretation stages to allow for a more comprehensive and rich understanding of contextual factors and a more complex evaluation of the issue. Previously, the MMR approach has been widely used in transportation research in the studies that combined qualitative data describing study participants' behaviors and preferences with quantitative survey and transportation mobility data (21-23).

**RESULTS**
**Survey Results Report**

The majority of survey participants said that their usual transportation mode is either driving alone or with other members of their household (87%). Among other transportation modes, survey participants reported using a bus or van service (6%), carpooling with friends or co-workers (4%), and walking (4%). The majority of participants (56%) reported having one car in their household while 27% of household reported having two cars in the household. 10% of respondents reported not having any cars in the household and 7% of respondents reported having three or more cars in the household.

Among public transportation options available in Pickens County, 71% of respondents reported using H.E.L.P. Inc. which is an origin to destination demand response and contract transportation agency serving with the goal of providing safe, reliable, affordable transportation for the citizens of Pickens County (6). Buses and 14% said they were using Church Transport to get around. Over one third of respondents (36%) said they had to miss a doctor's appointment because they did not have a way of getting there. The majority of participants said that when they need to go somewhere, they would always (43%) or sometimes (29%) need to plan ahead in order to be able to go. A third of respondents (33%) said that trying to get around town is a source of stress and anxiety for them with 16% of respondents sharing that they do not have a reliable means of getting around whenever they want.

When asked about their use of the H.E.L.P. bus, 32% of respondents shared that they have used H.E.L.P. bus in the past while 68% said they have never taken the H.E.L.P. bus. Among those respondents that have used the H.E.L.P. bus in the past, a little less than a third of respondents (31%) said they use it daily. 21% of respondents use it 1-2 times a week, 3% of respondents use it 1-2 times a month and 45% of respondents use it a few times a year. When it comes to the primary destination when using the H.E.L.P. bus, the most common use of the H.E.L.P. bus was for recreational purposes (27%), 20% of respondents stated they used the bus for other purposes, 18% said they used the H.E.L.P. bus to travel to work and 18% said they used the H.E.L.P. bus to travel to doctor's appointment. For 31% of respondents, the primary reason for selecting the H.E.L.P. bus over other ways to get around was the quality of service. Cost and the lack of other transportation options each accounted for 25% of responses. 19% of respondents said they chose the H.E.L.P. bus because of the ease of use. Among those respondents that choose not to use the H.E.L.P. bus, 50% stated that they prefer to drive, 10% said that "the bus is not for people like me." Other reasons why respondents chose not to use the H.E.L.P. bus included the fact that one has to schedule rides a day in advance, the need to use phone to schedule, the times of service and places serviced, and other.

**Focus Group Report**

The focus group discussions that were conducted in Gordo, Aliceville, Reform, and Carrollton generated some common themes among respondents as well as some responses that varied by location. The numbers of participants who took part in the focus group discussions in each of the focus groups locations are displayed in Table 2. Table 2-9 provide summaries of participants' responses for each of the two focus groups helps in Gordo, Aliceville, Reform, and Carrollton.

**Table 2. Number of participants that took place in each of the administered focus group discussions.**

|  | Gordo | Aliceville | Reform | Carrollton |
|---|---|---|---|---|
| Group 1 | 6 | 9 | 9 | 7 |
| Group 2 | 3 | 8 | 11 | 10 |
| Total | 9 | 17 | 20 | 17 |

**Table 3. Summaries of focus groups' responses regarding their means for getting around**.

|  | Gordo | Aliceville | Reform | Carrollton |
|---|---|---|---|---|
| Group 1 | Most participants drive, some of them rely on rides from family and relatives | Some of the participants drive their own cars; others use the H.E.L.P. bus to get around in town. | Some of the participants have cars, some use the H.E.L.P. bus, and some rely on friends to give a ride. Sometimes the participants are frustrated that they have to rely on other people to get around. | Most of the participants pay others to get a ride. The respondents feel the overall price of a ride is high because they are not only paying for gas, but also for the driver's time and the opportunity to use their vehicle. |
| Group 2 | Most of the participants get rides from family or neighbors or friends. Some of them walk, and some use the H.E.L.P. bus and church bus. | Most of the participants drive themselves. Some of them get rides from families. One participant stated that he travels by hitchhiking. | In this focus group, all participants either drive or ride with the family. All of them have a personal vehicle. | Some of the participants drive or have their own vehicle. Others either travel with their family or pay someone to take them. Some of them also use the H.E.L.P. bus to get around. |

From focus groups' responses, it is clear that most participants either use their personal vehicles or rely on others (mostly friends or family) for rides. Additionally, some of the participants mentioned using the H.E.L.P. bus and church bus to get around. Some of the participants mentioned that they find getting around to be a frustrating experience because of the lack of independence when it comes to travel and the overall cost of travel.

**Table 4. Summaries of focus groups' responses regarding planning ahead for getting around.**

| | Gordo | Aliceville | Reform | Carrollton |
|---|---|---|---|---|
| Group 1 | Those participants that do not have cars shared that they have to plan for their travel ahead of time unless it is a sudden emergency. | For most participants, getting around is not stressful but they feel that they must plan ahead to be able to go somewhere. Some of the participants said they drive locally and rely on others when they need to travel outside of their county of residence, especially in large urban areas like Birmingham. They do not feel safe around heavy traffic and sometimes they do not feel physically fit enough to drive long distances. | Most of the people need to plan before if they plan to go somewhere. Most of them use the H.E.L.P. bus. Therefore, they plan as per their schedule, especially for going to distant places like Tuscaloosa or going to the doctor. | Most of the participants shared they need to plan ahead and call someone beforehand to get around. |
| Group 2 | Participants said they need to plan ahead to be able to get around when they are traveling outside the county (especially in Birmingham). | All of the participants said they had to plan ahead to be able to get around regardless of them having a car or not. Some of them said they have to plan days beforehand. | Participants shared that most of the time, when they need to travel for a medical appointment, they need to plan ahead of time, especially if they have to go outside of their county of residence (e.g. Tuscaloosa, Birmingham, or Columbus). | All participants agreed with the statement that they have to plan ahead for getting around. They have to make certain that other people are available to give them a ride. |

Generally, the participants of focus group discussions shared that they had to plan ahead when traveling. In most cases, planning around was especially important for those participants that had to rely on others for rides and those that had to travel outside of their county of residence. Importantly, the participants of the Aliceville Focus Group 1 discussion mentioned that they felt safe using their own vehicle in their area of residence but preferred asking for a ride when it comes to traveling to larger urban areas such as Birmingham due to heavier traffic and the need to travel longer distances.

**Table 5. Summaries of focus groups' responses regarding the access to healthcare.**

|  | Gordo | Aliceville | Reform | Carrollton |
|---|---|---|---|---|
| Group 1 | Participants shared that traveling to medical appointments required planning ahead especially if the location of doctor's office was outside of county. | Some respondents said they relied on the rides from others when traveling to medical appointments. Even though planning ahead may be necessary when relying on the rides from others, planning ahead for medical travel may be challenging in case of next-day and emergency medical visits. | Participants shared they used the H.E.L.P. bus to travel to medical appointments, even outside of their county of residence. Additionally, respondents said they sometimes tried scheduling all of their appointments in a single day doctor's place frequently. If a family member or friend was giving a ride, when scheduling the appointment, it was important to ensure that the appointment time fits the driver's schedule. | All participants stated that at some point, they had to miss their doctor's appointment in the county and outside the county because they did not have a way of traveling to the appointment. |
| Group 2 | Some of the participants mentioned they had to miss their appointments because they did not have a car. | Respondents shared that even though some of the medical specialists did not fine patients for missing their appointments, most participants avoided missing their appointments and tried rescheduling their appointments if there was no way of getting there. | Some of the participants shared that they had to reschedule their doctor's appointments because they did not have a way to get there. Some of the appointments required a 72-hour notice prior the appointment in order to reschedule; otherwise, patients were responsible for additional charges. | Participants shared that they sometimes had to miss medical appointments both inside and outside of the county because they did not have a way of traveling to the appointment. |

Most participants shared that sometimes they had to miss their medical appointments because they did not have a way of traveling to the appointment. Most participants said that traveling to a medical appointment required planning ahead, especially if respondents had to rely on someone to give them a ride and if the appointment location was outside of their county of residence.

**Table 6. Summaries of focus groups' responses regarding the price of gas.**

|  | Gordo | Aliceville | Reform | Carrollton |
|---|---|---|---|---|
| Group 1 | Participants shared that the price of gas was important to them. | Most of the participants agreed that the price of gas affected their travel decisions. Obtaining medication at a lower price required traveling longer distances to Medicaid facilities (located in Tuscaloosa and Birmingham). The need to travel longer distances and increasing gas prices created challenges for their budget. | Most participants agreed that the price of gas was important to them. | Most participants said that the price of gas affected their travel choices. |
| Group 2 | The price of gas affected most participants. Participants believed that sometimes they had to pay more for travel when getting a ride from their family or friends than they would have if they were traveling in their personal vehicle. | All participants agreed that the price of gas affected their travel-related decision-making. | Participants stated that the high price of gas could be an extra burden on top of their regular vehicle-related expenses. | The price of gas was important to all participants regardless whether or not they owned a personal vehicle. |

Most participants agreed that the price of gas was important to them. The increasing price of gas was perceived to be a larger burden if participants had to travel longer distances or if they had to rely on others for a ride. Respondents believed that they had to pay extra when compensating others rides.

**Table 7. Summaries of focus groups' responses regarding stress in their experiences with getting around.**

|  | Gordo | Aliceville | Reform | Carrollton |
|---|---|---|---|---|
| Group 1 | Participants reported that they felt stressed about traveling outside of | Most participants did not feel stressed about getting around in town. One of the participants shared | Participants shared that they felt stressed when they needed to get a ride from others, especially if they were | Participants stated that getting around town was most stressful to them if they did not have |

| | | their county of residence | that when his car broke, he felt really stressed because he was not able to access the places where he needed to travel. | traveling out of town or for a medical visit. Participants shared that if a medical visit was going overtime, this sometimes caused additional stress because it affected the schedules of those individuals that were giving the patients a ride. | money for gas, or they could not find a ride. Traveling outside of their county of residence for a doctor's visit was especially stressful for them. |
| Group 2 | [absent from the report] | Some participants shared that they felt stressed while getting around while others did not feel stressed. Participants felt more stressed when they had to find a ride. | Most participants felt stressed about getting around because in the case of heavy traffic, road construction works and reckless drivers on the road. | Several participants reported feeling stressed about getting around because of their physical health. |

Participants shared that they experienced more stress when they needed to travel outside of their county of residence and when they had to rely on a ride from others. Additionally, participants experienced stress when it was challenging for them to afford a ride or when they were not able to use a personal vehicle because of vehicle breakdowns. Additionally, some of the reported causes of stress included difficult road conditions such as heavy traffic, construction works, and the presence of reckless and distracted drivers on the road as well as one's perceived lack of fitness to independently operate a personal vehicle.

**Table 8. Summaries of focus groups' responses regarding their experience with the H.E.L.P. bus**

| | Gordo | Aliceville | Reform | Carrollton |
| Group 1 | None of the participants used the H.E.L.P. bus because they thought it is a service for a certain | Some of the participants like using the H.E.L.P. bus because it creates an opportunity for | Most of them have positive experiences with the H.E.L.P. bus. They used the H.E.L.P. bus for a variety of purposes, | One of the participants used the H.E.L.P. bus to travel to Bevill State Community College but he stopped |

| | | | |
|---|---|---|---|
| | category of population (such as people with disabilities) for which respondents may not qualify. | them to socialize with other passengers. Some of them were disappointed with the lack of the H.E.L.P. bus' schedule reliability. One participant mentioned that the H.E.L.P. bus is helping him get around locally but not when it comes to traveling longer distances (such as to Tuscaloosa or Birmingham). Those participants that use the H.E.L.P. bus to travel to medical appointment suggested that the H.E.L.P. bus service area should cover Birmingham and Tuscaloosa since most of medical services providers are located there. | like going to doctors, groceries, banks, etc. The participants found the H.E.L.P. bus service to be very convenient. | going there. Other participants used the H.E.L.P. bus when they had to go to medical appointments. A few participants reported that, they have never used the H.E.L.P. bus, but some of their family members have used it. |
| Group 2 | None of the participants used the H.E.L.P. bus, but some participants reported that their family members used the service. One of the participants said they had a negative experience of scheduling with the H.E.L.P. bus because they were dissatisfied with H.E.L.P. bus' phone customer service. | Only one participant in the group said they used the H.E.L.P. bus. That participant used the bus service to travel to therapy in Carrolton. Participants' suggestions regarding improving the H.E.L.P. bus service included expanding the bus service to areas outside the county (e.g. to Tuscaloosa | The participants are mostly happy with the H.E.L.P. bus service. Even those participants that said they usually rely on personal vehicle as their main transportation mode, said they have used the H.E.L.P. bus in the past. Most participants have had a good experience with the | The participants said that them, as well as their family members, used the H.E.L.P. bus mostly for travel of their county of residence or to go to the senior center. |

| | | County) even if adding longer service routes may demand increasing ride cost on those routes. | H.E.L.P. bus drivers. | |
|---|---|---|---|---|

In most focus groups, participants seemed to be familiar with the H.E.L.P. bus service as they reported that either them or their family members have used the bus service in the past. Interestingly, none of the participants from the Focus Group 1 in Gordo said they had used the H.E.L.P. bus service previously because they were not certain whether or not they would qualify to use the service. This finding highlights the importance of outreach and H.E.L.P. bus service advertisement to ensure that residents are aware of all transportation options available to them. Generally, those participants that have used the H.E.L.P. bus service previously have had a positive experience using the service. Respondents used the bus service for different purposes including medical visits, grocery trips, and others. Some of the points of dissatisfaction with the bus service included the lack of the bus service punctuality and the low quality of the phone customer service. In two of the focus groups, respondents mentioned that it would be helpful to expand the H.E.L.P. bus service and include longer-distance trips such as Tuscaloosa County and Birmingham.

**Table 9. Summaries of focus groups' responses regarding their travel option improvement opinions**

| | Gordo | Aliceville | Reform | Carrollton |
|---|---|---|---|---|
| Group 1 | Participants believed that improving communication in communities and developing personal support networks would be helpful as it would allow finding rides when needed. | Participants said it would be helpful to have another bus service available which, unlike the H.E.L.P. bus operates on a fixed schedule. They suggested that it might be helpful to have a community group that would give rides in an organized way to members of the community. Another suggestion was for churches to give out grants to provide transportation | Respondents emphasized the importance of improving the conditions of local roads such as repairing potholes in pavements. | Participants emphasized the importance of improving the condition of the roads. |
| | | | | services to residents. |
| Group 2 | The participants said that it would be helpful if taxi, Uber, and Lyft services could be expanded to rural areas such as Gordo. Respondents emphasized that they would like to have local transportation that will serve them inside and outside the county. | As a travel option improvement option, the participants demanded more road lights in Aliceville. They also mentioned about dips and potholes on roads that need to be fixed. | The participants are really disappointed with the present condition of their roads. For Tuscaloosa, since they are always fixing roads, it's a hassle to get there. Also, the ride charge is more because of that. The roads are also dangerous for their children who are going to school. | The participants mostly suggested for providing street lights, fixing ditches, and potholes on roads. They are really frustrated that big trucks are coming to the neighborhoods and ruining the road pavements. Because of this, their kids can't play scooters on roads; they play videogames instead. |

**DISCUSSION**

The results of the study are consistent with previous research that has showed that having access to a personal vehicle is perceived to be the most convenient way to get around in rural communities. Individuals that participated in the survey conducted as a part of this research study reported that relying on others for a ride or using demand-responsive public transit options such as H.E.L.P. bus may be a stressful and frustrating experience because of the lack of independence, unreliability, and high associated transportation costs. Additionally, participants shared that the lack of reliable transportation options has caused them to miss or delay medical appointments. The fact that study participants emphasized the importance of planning for travel ahead of time, especially in the case of long-distance trips and travel to medical appointments reveals the need for more transportation options that would be accessible, affordable, reliable and punctual. Given that some medical conditions may require unplanned next-day or emergency travel to a medical facility, not having transportation options that would be readily available in such situations poses a serious threat to communities' public health.

This research study highlighted some of the transportation-related challenges in rural communities as well as the need for more available transportation options in rural areas. One of the findings of the study was that most survey respondents found their transportation experiences with the local demand-response bus service to be overall positive and helpful. This suggests that one of possible ways to improve mobility in rural areas and address the gap in rural public transportation services would be to expand demand-response transportation (DRT) services and transportation network companies (TNC) services to rural areas because currently, the latter are concentrated predominantly in urban areas (24).

**CONCLUSION**

Through the adoption of a mixed methods research approach, this study aimed to examine the transportation needs of the residents of Pickens County, Alabama. Particularly, the study looked at transportation-related challenges experienced by the residents of Pickens County, and how such challenges may affect their overall quality of life including their access to healthcare, employment, and essential services. Additionally, the study elicited and analyzed Pickens County residents' opinions regarding the possible ways to improve transportation services in the county area.

The study revealed that amongst participants who did not own a vehicle, the majority relied upon either friends/family members or had to pay someone to obtain a ride. About 3% of respondents shared that they resorted to walking whereas biking was even less common. Most significantly, it was found that only a tiny fraction of respondents reported using the H.E.L.P. bus as a way to get around.

Many of those respondents that did not own a personal vehicle shared that traveling was often a stressful experience for them. Additionally, since medical visits often required long-distance trips, participants shared that the travel to medical appointments was also often a source of stress to them. Even amongst those with vehicle ownership, heavy traffic, road works and reckless drivers on the roads causes stress. Furthermore, vehicle breakdown is also reported to induce stress due to the loss of mobility.

When it came to actually making the trips, most participants reported that they had to plan ahead for their journeys, irrespective of whether or not they owned a vehicle. People who were reliant upon others for transportation had to ensure that the person they would seek rides from were available at the correct time. Planning also involved accounting for the cost of the ride in the form of gas money. Even for the participants having their own cars, it was reported by some that they felt discomfort driving in the heavy traffic outside of the county and thus needed to plan rides with other people.

Additionally, this study highlighted the need to expand the outreach and increase the awareness about the public transit options that are available in the area. For example, participants of the first focus group from Gordo reported that they never used H.E.L.P. bus services due to the misconception that its services were only reserved for handicapped people. The discussions also circulated around the positive and negative experiences people had when they used the H.E.L.P. bus services. Respondents liked that the H.E.L.P. bus was affordable for the passengers, especially the senior citizens. The negative experiences were all tied to the lack of service flexibility.

Finally, the study obtained participants' opinions about the improvement of travel options and shuttle services. Participants mentioned about repairing existing roads, introducing more streetlights and restricting heavy commercial vehicles that can ruin the pavement. Participants unanimously voiced the need for a proper regular bus service in Pickens County that can transport them to distant locations. They also suggested that a better communication within the community can help in the efficient management of ride sharing. Participants from Gordo expressed interest in expanding private ride hailing services like Uber and Lyft to Pickens

County. As Pickens County had no private transit services like Greyhound buses, the participants believed that the HELP bus authorities should obtain additional funding in order to expand their operations throughout the whole county by introducing more buses, drivers and increasing service coverage. At the same time, they wanted the HELP buses to reach more distant locations like Tuscaloosa, Birmingham and Columbus. Even within the current service area, the HELP bus services are inadequately publicized, and resident felt that the authorities should expend more efforts towards advertising this service. The bus schedules should be made more convenient for the riders. Some suggested that there should be a warning system to alert the passengers when the bus is near their location to prevent them from missing the bus. Lastly, the accessibility of the buses should be improved by incorporating ramps, wheelchairs etc.

It is expected that the findings of this transportation need assessment will help shed light on the transportation issues that affect the residents of Pickens County. Alongside this, the participants' suggestions will be beneficial for improved design of transit services in Pickens County as well as improved operations of the H.E.L.P. bus service. Future works can expand upon this study by replicating the methods in the context of other rural communities throughout the state in order to examine the similarities as well as disparities between the communities.

## ACKNOWLEDGMENTS

This work is supported by the Alabama Transportation Institute at The University of Alabama.

## AUTHOR CONTRIBUTIONS
The authors confirm contribution to the paper as follows:
study conception and design: Riffat Islam, Steven Jones; data collection: Riffat Islam, Steven Jones; analysis and interpretation of results: Riffat Islam, Olga Bredikhina, Khadiza Tul Jannat; draft manuscript preparation: Olga Bredikhina, Muhammad Sami Irfan, Khadiza Tul Jannat, and Riffat Islam. All authors reviewed the results and approved the final version of the manuscript.

**REFERENCES**

1. Brown, D. M. Public Transportation on the Move in Rural America. Economic Research Service, U.S. Department of Agriculture, 2004

2. McDaniels, B.W., Harley, D.A., Beach, D.T. Transportation, Accessibility, and Accommodation in Rural Communities. In: *Disability and Vocational Rehabilitation in Rural Settings* (Harley, D., Ysasi, N., Bishop, M., Fleming, A., ed.), Cham: Springer, 2018, p. 43–57. https://doi.org/10.1007/978-3-319-64786-9_3

*3.* Henning-Smith, C., Evenson, A., Corbett, A., Kozhimannil, K., & Moscovice, I. Rural Transportation: Challenges and Opportunities. University of Minnesota Rural Health Research Center, 2017. http://rhrc.umn.edu/wp-content/files_mf/1518734252UMRHRCTransportationChallenges.pdf

4. American Hospital Association. Social Determinants of Health Series: Transportation and the Role of Hospitals. Chicago, IL: Health Research & Educational Trust, 2017. https://www.aha.org/system/files/hpoe/Reports-HPOE/2017/sdoh-transportation-role-of-hospitals.pdf

5. Massey, J., Beezer, D., & Stauber, A. Community Transportation Needs Assessment: Process, Results, and Lessons Learned. TransForm, 2020. https://www.transformca.org/transform-report/community-transportation-needs-assessment

6. Hamilton, B. H.E.L.P., Inc. to Begin Transportation to Tuscaloosa. *Pickens County Herald*, 2019. https://pcherald.com/node/2100

7. Federal Highway Administration. Statewide Transportation Planning. Accessed 30 July 2022. https://www.fhwa.dot.gov/planning%20/processes/statewide/related/highway_functional_classifications/section06.cfm

8. Kidder, B. The Challenges of Rural Transportation. Western Rural Development Center, Utah State University, 2006

9. Ellis, E. H., & McCollom, B. E. Guidebook for Rural Demand-Response Transportation: Measuring, Assessing, and Improving Performance, Vol. 136, 2009. Transportation Research Board.

10. Council of Economic Advisers. Strengthening the Rural Economy. Executive Office of the President, 2010. https://www.agri-pulse.com/ext/resources/pdfs/r/u/r/1/0/RuralAmericaRpt27Apr10.pdf

11. Hartman, R. and Weierbach, F. Elder health in rural America. National Rural Health Association. Policy Brief; February 2013. https://www.ruralhealth.us/getattachment/Advocate/Policy-Documents/ElderHealthinRuralAmericaFeb2013.pdf.aspx

12. Criden, M. The Stranded Poor: Recognizing the Importance of Public Transportation for Low-Income Households. National Association for State Community Services Programs, 2008.

13. Symens Smith, A. and Trevelyan, E. In Some States, More than Half of Older Residents Live in Rural Areas. United States Census Bureau, 2019. https://www.census.gov/library/stories/2019/10/older-population-in-rural-america.html

14. Arcury, T. A., Preisser, J. S., Gesler, W. M., & Powers, J. M. Access to Transportation and Health Care Utilization in a Rural Region. *The Journal of Rural Health*, *21*(1), 31-38, 2005.

15. Fletcher, C. N., Garasky, S. B., Jensen, H. H., & Nielsen, R. B. Transportation Access: A Key Employment Barrier for Rural Low-Income Families. *Journal of Poverty*, *14*(2), 123-144, 2010.

16. Morgan, D. L., and Krueger, R. A. When to Use Focus Groups and Why. *Successful Focus Groups: Advancing the State of the Art*, *1*, 3-19, 1993.

17. O'Cathain, A. Mixed Methods Research in the Health Sciences: A Quiet Revolution. *Journal of Mixed Methods Research*. 2009; 3(1):3-6. doi:10.1177/1558689808326272

18. Mertler, C. A. *Introduction to educational research*. Sage publications, 2021.

19. Taguchi, N. (2018). Description and Explanation of Pragmatic Development: Quantitative, Qualitative, and Mixed Methods Research. *System*, *75*, 23-32, 2018.

20. Schoonenboom, J., & Johnson, R. B. How to Construct a Mixed Methods Research Design. *Kolner Zeitschrift fur Soziologie und Sozialpsychologie*, *69*(Suppl 2), 107–13, 2017. https://doi.org/10.1007/s11577-017-0454-1

21. Lucas, K., Philips, I., & Verlinghieri, E. A Mixed Methods Approach to the Social Assessment of Transport Infrastructure Projects. *Transportation*, *49*(1), 271-291, 2022.

22. Shay, E., Combs, T. S., Findley, D., Kolosna, C., Madeley, M., & Salvesen, D. Identifying Transportation Disadvantage: Mixed-Methods Analysis Combining GIS Mapping with Qualitative Data. *Transport Policy*, *48*, 129-138, 2016.

23. Fang, X., Cao, C., Chen, Z., Chen, W., Ni, L., Ji, Z., & Gan, J. Using Mixed Methods to Design Service Quality Evaluation Indicator System of Railway Container Multimodal Transport. *Science Progress*, *103*(1), 0036850419890491, 2020.

24. Feigon, S., & Murphy, C. (2018). Broadening understanding of the interplay among public transit, shared mobility, and personal automobiles (No. Project J-11/Task 25)